\documentclass[12pt,preprint]{aastex62}
\usepackage{graphicx}
\usepackage{longtable}
\usepackage{multirow}
\usepackage{url}
\usepackage{amsmath}
\usepackage{lipsum}
\usepackage{rotating}
\newcommand{\dd}{\ensuremath{\text{d}}}
\begin{document}

 \title{A universal energy distribution for FRB 121102}


\author[0000-0003-4157-7714]{F. Y. Wang}
\affiliation{School of Astronomy and Space Science, Nanjing University, Nanjing 210093, China}
\affiliation{Key Laboratory of Modern Astronomy and Astrophysics (Nanjing University), Ministry of Education, Nanjing 210093, China}

\author[0000-0001-6545-4802]{G. Q. Zhang}
\affiliation{School of Astronomy and Space Science, Nanjing University, Nanjing 210093, China}

\correspondingauthor{F. Y. Wang}
\email{fayinwang@nju.edu.cn}

\begin{abstract}
Fast radio bursts (FRBs) are milliseconds radio transients with
large dispersion measures (DMs). An outstanding question is the
relation between repeating FRBs and those with a single burst. In
this paper, we study the energy distribution of the repeating FRB
121102. For a power-law distribution of energy $dN/dE\propto
E^{-\alpha_E}$, we show that the value of $\alpha_E$ is in a narrow
range $1.6-1.8$ for bursts observed by different telescopes at
different frequencies, which indicates a universal energy
distribution for FRB 121102. Interestingly, similar power-law index
of energy distribution for non-repeating FRBs observed by Parkes and
ASKAP is also found. However, if low-energy bursts below
completeness threshold of Arecibo are discarded for FRB 121102, the
slope could be up to 2.2. Implications of such a universal energy
distribution are discussed.
\end{abstract}

\section{Introduction}
Fast radio bursts (FRBs) are radio transients with extreme
brightness temperatures that show large dispersion measurement (DM)
\citep{Lorimer2007Sci...318..777L,
Thornton2013Sci...341...53T,Petroff2016PASA...33...45P}. At present,
FRB 121102 and FRB 180814.J0422+73 show repeating bursts
\citep{Spitler2014ApJ...790..101S,Spitler2016Natur.531..202S,Amiri2019a}.
FRB 121102 has been localized to a z=0.19 galaxy
\citep{Chatterjee2017Natur.541...58C}. A large sample of FRBs with
redshifts can be used as potential cosmological probes
\citep{Deng2014ApJ...783L..35D,Gao2014,
McQuinn2014,Zheng2014,Zhou2014,Wei2015,Yu2017A&A...606A...3Y,Yang2016,Macquart2018MNRAS.480.4211M,
Li2018NatCo...9.3833L,Wang2018,Walters2018,Li2019}. Before its
cosmological implications, its progenitor should be known. Many
progenitor models have been proposed for FRBs. However, the physical
origin of FRBs is still mystery at present
\citep{Pen2018NatAs...2..842P}. A fundamental issue is relation
between repeating FRBs and non-repeating ones
\citep{Caleb18,Palaniswamy18}.

Many works have been carried out to statistically study FRBs from
Parkes and ASKAP samples, which can give important constraints on
their progenitors \citep{Katz2016,Lu2016MNRAS.461L.122L,Li2017RAA....17....6L,
Macquart2018MNRAS.480.4211M,Luo2018MNRAS.481.2320L,Cao2018ApJ...858...89C,
Lu2019arXiv190300014L}.
However, there is no redshift
information for the FRBs in Parkes and ASKAP samples. The energy or luminosity
derived from pseudo redshift is not reliable. Moreover, there are
some selection effects for the two samples
\citep{Keane2015MNRAS.447.2852K,Ravi2019MNRAS.482.1966R}.

In this paper, we focus on the repeating FRB 121102, which is
discovered by Arecibo telescope at 1.4 GHz
\citep{Spitler2014ApJ...790..101S}. It has also been observed by
Green Bank Telescope (GBT) at 2 GHz
\citep{Scholz2016ApJ...833..177S,Scholz2017ApJ...846...80S}, VLA at
3.5 GHz \citep{Chatterjee2017Natur.541...58C,Law2017}, GBT at 4-8
GHz \citep{Gajjar2018ApJ...863....2G,Zhang2018ApJ...866..149Z},
Arecibo telescope at 4.1-4.9 GHz \citep{Michilli2018Natur.553..182M}
and Effelsberg radio telescope at 4.85 GHz
\citep{Spitler2018ApJ...863..150S}. There are more than 200 bursts
from FRB 121102 at frequencies ranging from 1 to 8 GHz. An
outstanding question appears, do these bursts observed by different
telescopes at different frequencies follow a similar distribution?
\citet{Wang2017} studied the frequency distributions of peak flux,
fluence, duration and waiting time for bursts observed by Arecibo
telescope at 1.4 GHz, and found these distributions are similar to
those of soft gamma-ray repeaters (SGRs). The energy distribution is
$dN/dE\propto E^{-1.8}$ \citep{Wang2017}. Similar energy
distributions $dN/dE\propto E^{-1.7}$ of the VLA, Arecibo, and GBT
bursts are found by \citet{Law2017}. However, a much steeper value
of $\alpha_E=2.8$ is derived by \citet{Gourdji2019arXiv190302249G}
using 41 bursts observed Arecibo telescope at 1.4 GHz. Therefore,
whether a universal energy distribution exists for FRB 121102 is
controversial. In this paper, we study the energy distribution of
FRB 121102 observed by Arecibo telescope, GBT and VLA at different
frequencies.

The paper is organized as follows. The burst samples are listed in section 2. In
section 3, we give the energy distributions. Finally, conclusion and
discussions are given in section 4. Throughout this
paper, we adopt a flat $\Lambda$CDM model with $\Omega_M$=0.27 and
$H_0 = 70$ km s$^{-1}$ Mpc$^{-1}$.

\section{FRB samples}
We collect the bursts of FRB 121102 by different radio
telescopes at different frequencies, including the observation of
VLA at 3 GHz \citep{Chatterjee2017Natur.541...58C}, the observation
of Arecibo at 1.4 GHz \citep{Spitler2016Natur.531..202S}, the
observation of Arecibo at 4.5 GHz
\citep{Michilli2018Natur.553..182M}, the observation of GBT at 4-8
GHz \citep{Zhang2018ApJ...866..149Z}, the observation of GBT at 2
GHz \citep{Scholz2016ApJ...833..177S,Scholz2017ApJ...846...80S} and
the recent observation of Arecibo at 1.4 GHz
\citep{Gourdji2019arXiv190302249G}. For observation, burst energy can be calculated through
\begin{equation}
    \label{eq:ene}
    E = 4 \pi d_L^2 F \Delta \nu
\end{equation}
where $d_L$ is luminosity distance, $F$ is burst fluence and
$\Delta \nu$ is the bandwidth of the observation. For the
observation of VLA at 3 GHz by
\citet{Chatterjee2017Natur.541...58C}, we adopt the energies given
by \citet{Law2017}. We list the observation time
(MJD), fluence, central frequency, bandwidth, duration and energy for different samples in tables \ref{tab:law} - \ref{tab:zhang}. In
tables \ref{tab:gourdji} and \ref{tab:zhang}, the central frequency
and the bandwidth are derived from the high frequency and low
frequency given by \citet{Gourdji2019arXiv190302249G}. For the
observation of \citet{Michilli2018Natur.553..182M}, we only consider
first 16 bursts in their paper. Besides, only bursts observed by GBT
at 2 GHz are considered in our analysis for the observation by
\citet{Scholz2016ApJ...833..177S, Scholz2017ApJ...846...80S}.
Using these data, we derive the energy function of FRB 121102 at
different frequencies.

\section{Results}
\begin{figure*}
        \centering
        \includegraphics[width=0.8\linewidth]{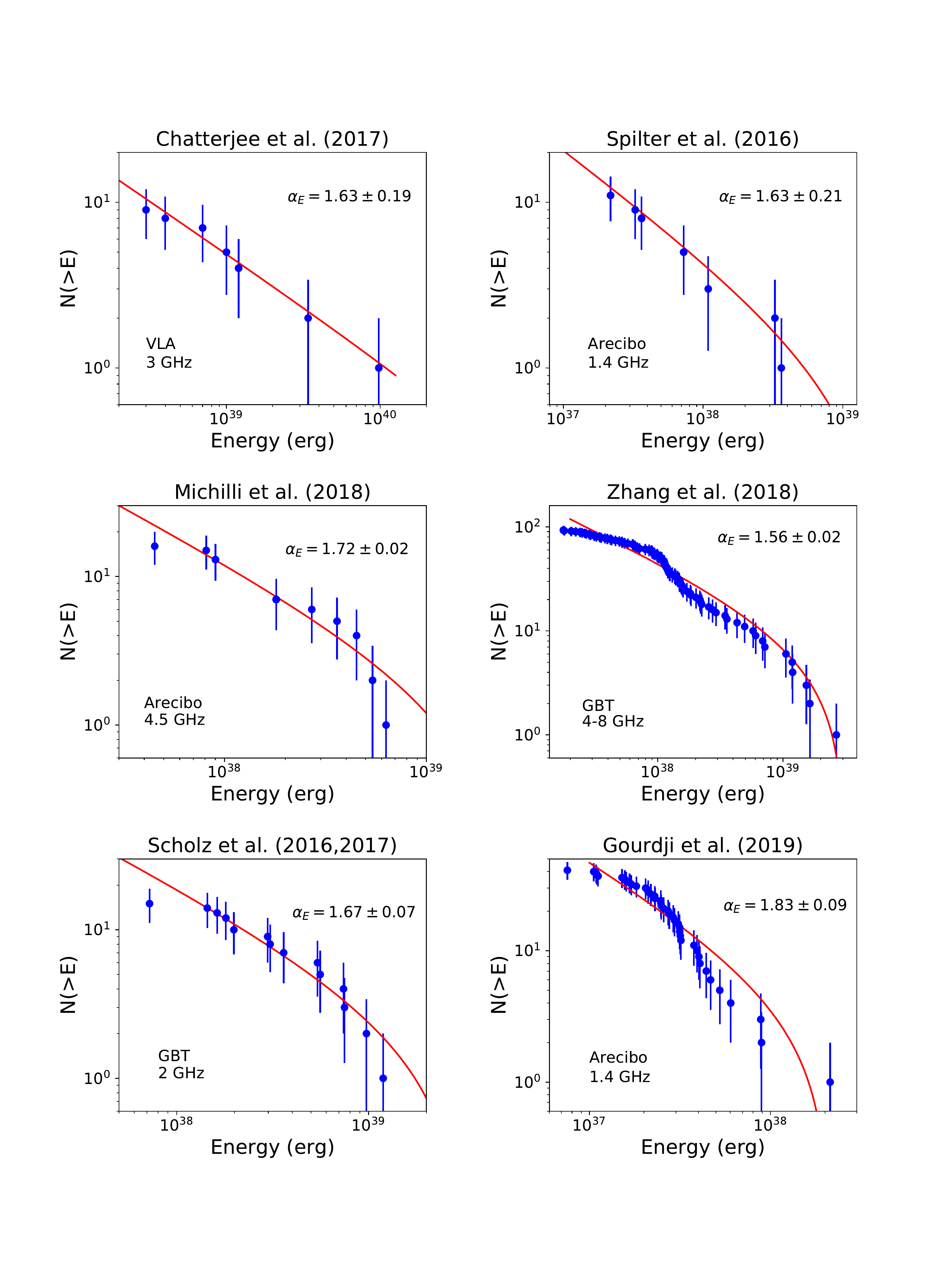}
        \caption{The cumulative energy distributions for different samples of FRB 121102. The observed frequencies
        and telescopes are shown.
        The best-fitting power-law index $\alpha_E$ for a power-law function with a high energy cut-off (equation (2))
        is in a narrow range $1.6<\alpha_E<1.8$.}
        \label{fig:frbE}
    \end{figure*}

For small samples, a cumulative distribution is often used, because
the small number of bursts is not sufficient to bin the data. A
cumulative distribution is defined by the integral of the total
number of events above a given value. It should be noted that we
must consider the deviation from ideal power-law distribution. There
are many effects that cause this deviation, such as the threshold of
telescope and a physical threshold of an instability
\citep{2015ApJ...814...19A}. Therefore, we adopt threshold power-law
distribution with high-energy cutoff to fit the cumulative
distribution, which is
\begin{equation}
\label{eq:Ndis}
N(>E)=A(E^{1-\alpha_E}-E_{\rm{max}}^{1-\alpha_E}),
\end{equation}
where $E_{\rm max}$ is the maximum energy of FRB and $\alpha_E$ is
the power-law index of differential distribution $dN/dE\propto
E^{-\alpha_E}$. Observationally, the high-energy cutoff $E_{\rm
max}$ is clearly shown in the samples of
\citet{Michilli2018Natur.553..182M},
\citet{Scholz2016ApJ...833..177S}, \citet{Zhang2018ApJ...866..149Z}
and \citet{Gourdji2019arXiv190302249G} (Figure \ref{fig:frbE}).
Theoretically, every astronomical phenomenon must have a maximal
energy. \citet{2015ApJ...814...19A} discussed some theoretical
reasons for the cutoff, such as contamination by an event-unrelated
background and truncation effects at the largest events due to a
finite system size. Evidences for the high-energy cutoff have been
found in cumulative distributions of solar flares and stellar flares
\citep{2015ApJ...814...19A}, gamma-ray bursts
\citep{Wang2013NatPh...9..465W} and soft-gamma repeaters
\citep{Prieskorn2012ApJ...755....1P}. \citet{Lu2019MNRAS} found the
maximal luminosity of FRBs under the assumption that FRBs are from
coherent curvature emission powered by the dissipation of magnetic
energy in the magnetosphere of neutron stars. In addition, due to
telescope sensitivity, observational data at low energy may be
incomplete, which has been illustrated by
\citet{Gourdji2019arXiv190302249G}. Therefore, the completeness
threshold at low energy must be considered. We try to fit the
cumulative distributions with power-law function in three cases: a
power-law function with high-energy cutoff, a power-law function
with low-energy threshold and high-energy cutoff, a power-law
function with low-energy threshold. The low-energy threshold can be
taken into account by omitting bursts below the observational
sensitivity. We calculate the sensitivity of VLA at 3 GHz according
to \citet{Law2017} and the sensitivity of GBT at 2 GHz according to
\citet{Scholz2016ApJ...833..177S}. We adopt $2 \times 10^{37} $ erg
as the sensitivity of Arecibo at 1.4 GHz, which has been used in
\citet{Gourdji2019arXiv190302249G}. Using the fluence limit given by
\citet{Zhang2018ApJ...866..149Z}, we calculate the energy limit with
the mean bandwidth. Only considering the data with signal-to-noise
larger than 5, the energy limit can be derived as $1.02\times
10^{38}$ erg for observation of Arecibo at 4.5
GHz\citep{Michilli2018Natur.553..182M}. With these sensitivities, we
fit the cumulative distributions with an ideal power-law function or
a power-law function with cutoff.

The best fitting parameters can be derived by minimizing
\begin{equation}\label{chi2}
\chi^{2}=\sum _{ i=1 }^ n  \frac{\left( N_{ob,i}-N(>E) \right) ^ 2}{\sigma_{ob,i}^2} ,
\end{equation}
where ${N}_{ob}$ is the number of observed bursts, $N(>E)$ is the
model-predicted number of bursts from eq.\ref{eq:Ndis} and
$\sigma_{ob}$ is the uncertainty of the cumulative distribution. The
expected uncertainty of the cumulative distribution is
\citep{2015ApJ...814...19A}
\begin{equation}
\sigma_{ob,i}=\sqrt{N_i}.
\end{equation}
We use an open source python package emcee
\citep{Foreman2013PASP..125..306F} to constrain the parameters
($A,\alpha_E$ and $E_{\rm max}$) through Markov chain Monte Carlo
method. As discussed by \citet{Aschwanden2011soca.book.....A} (in
page 204 of the book), in reality, there is always a largest event,
which causes a gradual steepening in the cumulative frequency
distribution, because of the missing contributions up this point.
This feature is quite important, because it leads to a significant
over-estimate of the power-law slope. The best way is to fit the
exact analytical function of the cumulative frequency distribution
function $N(>E)=A(E^{1-\alpha_E}-E_{\rm max}^{1-\alpha_E})$. In our
paper, for the cumulative distribution shows obvious steepening in
high-energy end, we consider the parameter $E_{\rm max}$ in the
fitting. Because the distribution of VLA bursts
\citep{Chatterjee2017Natur.541...58C} shows ideal power law, the
$E_{\rm max}$ is not included in the fitting.

\begin{figure*}
    \centering
    \includegraphics[width=0.8\linewidth]{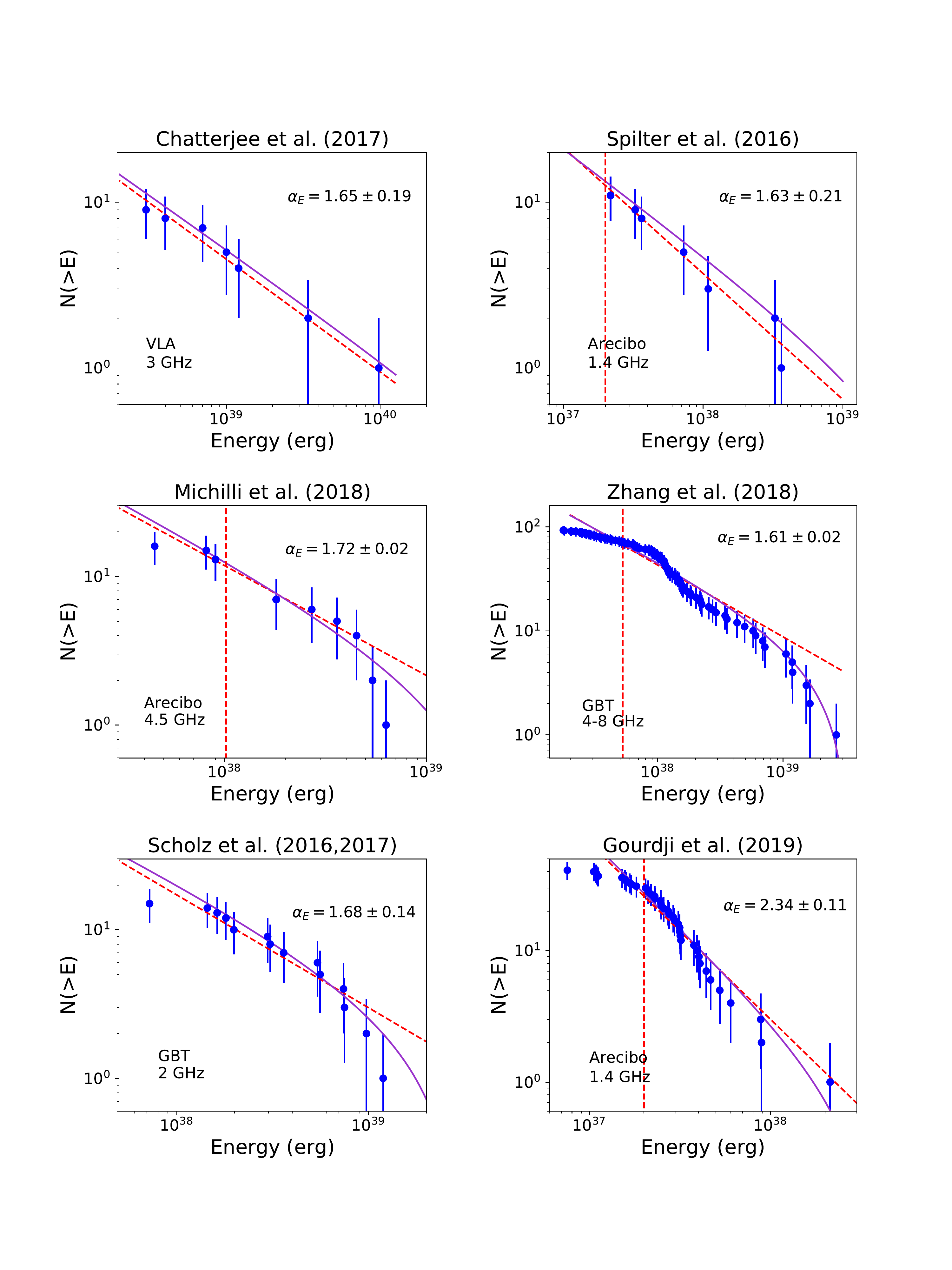}
\caption{The cumulative energy distributions for different samples of FRB 121102. The observed
frequencies and telescopes are shown. The purple solid line is the
best fitting by a power-law function with low-energy threshold and high-energy cutoff. The value of $\alpha_E$ is for purple solid lines. The red dashed line is the fit by a power-law
function with low-energy threshold. The low-energy thresholds are
shown as vertical red dashed lines.}
    \label{fig:frbEother}
\end{figure*}

\begin{figure*}
    \centering
    \includegraphics[width=0.8\linewidth]{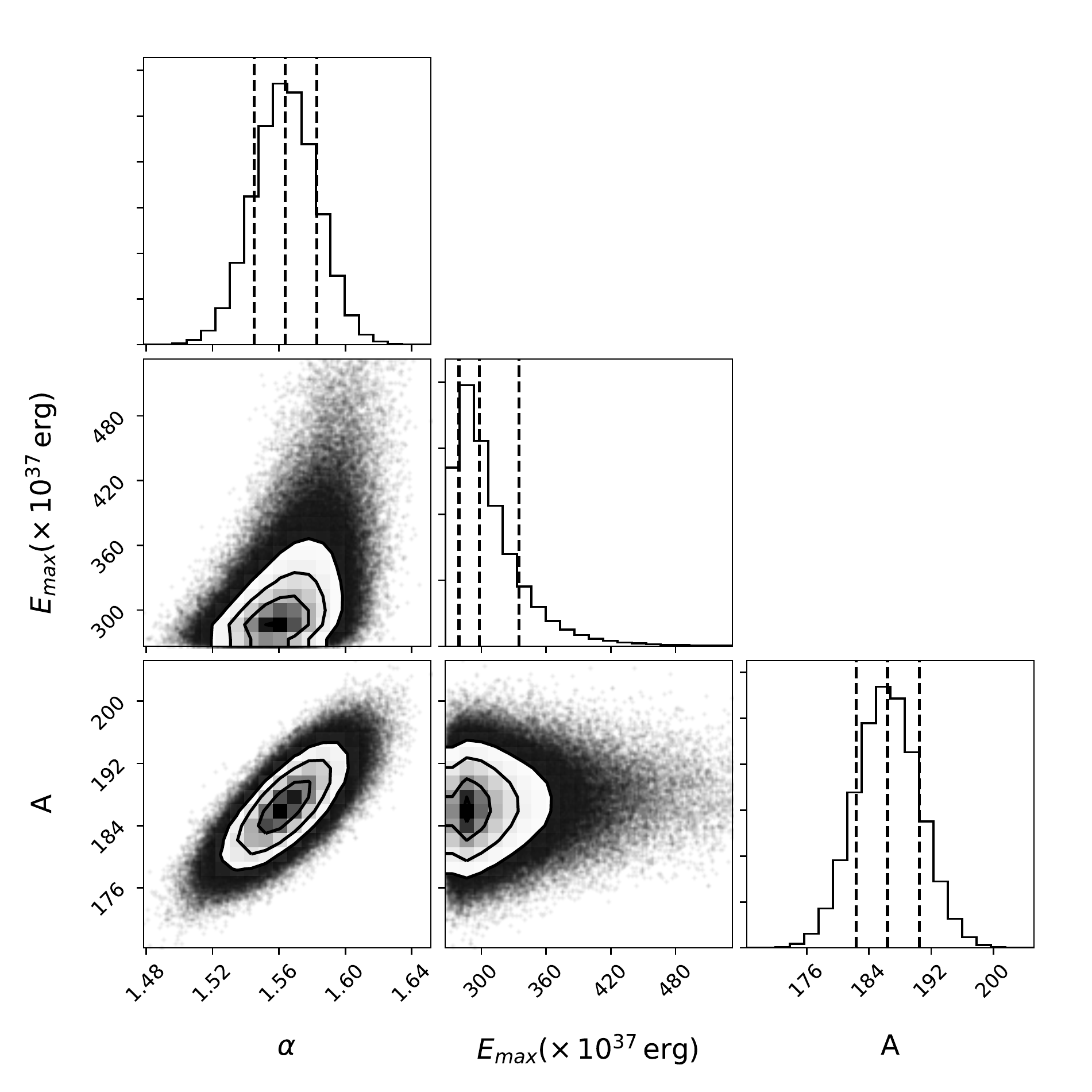}
    \caption{The corner plot of the sample for \cite{Zhang2018ApJ...866..149Z}.}
    \label{fig:corner}
\end{figure*}

Figure \ref{fig:frbE} shows the fitting results of cumulative
distributions for different samples with only high-energy cutoff.
The best-fitting power-law indices are $\alpha_E=1.63\pm0.19$,
$1.63\pm0.21$, $1.72\pm0.02$, $1.56\pm0.02$, $1.67\pm0.07$ and
$1.83\pm0.09$ for samples from
\citet{Chatterjee2017Natur.541...58C},
\citet{Spitler2016Natur.531..202S},
\citet{Michilli2018Natur.553..182M},
\citet{Zhang2018ApJ...866..149Z},
\citet{Scholz2016ApJ...833..177S,Scholz2017ApJ...846...80S} and
\citet{Gourdji2019arXiv190302249G}, respectively. The error
represents the $ 1\sigma $ confidence level. We also list all the
best-fitting parameters in Table \ref{tab:results}. In these fitting
results, the maximal energy is about $ 200 \sim 300 \times 10^{37} $
erg except the observations of VLA and the sample of
\citet{Gourdji2019arXiv190302249G}. The maximal energy is only $
(21\pm17) \times 10^{37} $ erg for the sample of
\citet{Gourdji2019arXiv190302249G}, which is smaller than others.
The reason is that the average energy of this sample is lower than
others. Besides, through the MCMC method, we also obtain the corner
plot of the fitting results. Figure \ref{fig:corner} shows the
corner plot for samples from \citet{Zhang2018ApJ...866..149Z}. It is
interesting that the value of $\alpha_E$ is in a narrow range, i.e.,
from 1.6 to 1.8. Therefore, a universal energy distribution for FRB
121102 is found. This result is consistent with those of
\citet{Wang2017} and \citet{Law2017}. In the other two cases, we
show the fitting results in Figure \ref{fig:frbEother} and Table
\ref{tab:Otherresults}. In Figure \ref{fig:frbEother}, the purple
solid line shows the best fitting with low-energy threshold and
high-energy cutoff. The red dashed line represents the best fitting
with low-energy threshold, which deviates the observational data at
high energies. Therefore, a power-law function can not fit the data
by omitting bursts below the observational sensitivity. The
high-energy cutoff must be considered. The low-energy sensitivity is
shown as vertical red dotted line. From Table
\ref{tab:Otherresults}, we can see that the value of $\alpha_E$ is
between 1.6 and 1.8, except for the sample of Gourdji et al. (2019).
A steeper value $\alpha_E=2.8$ is found by
\citet{Gourdji2019arXiv190302249G} using 41 bursts observed Arecibo
telescope at 1.4 GHz. This discrepancy may be caused by choosing
threshold energy. If these low-energy bursts are considered, the
value of $\alpha_E$ is around 1.8 (figure 5 of
\citet{Gourdji2019arXiv190302249G}), which is consistent with our
result. In addition, the Galactic scintillation also affect the
observed radio emission from impulsive radio sources at high
frequency \citep{Macquart2015MNRAS.451.3278M}. For example, the GBT
bursts are affected by scintillation at 4-8 GHz, and this
complicates calculation of the burst energies
\citep{Gajjar2018ApJ...863....2G,Hessels2018arXiv181110748H}.
Therefore, the fitting result for GBT bursts is not quite well,
which is shown in the panel 4 of figure \ref{fig:frbE}.

\section{Discussion and Conclusions}
In this paper, the energy distribution of the repeating FRB 121102
is studied using six samples observed at different frequencies. We
find a universal energy distribution for FRB 121102 with power-law
index $1.6<\alpha_E<1.8$. However, if low-energy bursts below
completeness threshold of Arecibo are discarded for FRB 121102, the
slope could be up to 2.3. Some of the implications of our results
are as follow.

First, we discuss the volumetric birth rate $R_{FRB}$ of the
repeating FRBs. If the life time of each repeater is $\tau$ years,
the volume density is $R_{\rm FRB}\tau$. Assuming the formation rate
of FRBs tracks the star formation rate
\citep{Hopkins2006ApJ...651..142H}, the volumetric birth rate
at redshift $z<1$ is $R_0(1 + z)^{3.28}$, where $R_0$ is the local
formation rate of FRBs. Considering each repeating FRB has
$r(>E_{\rm min})$ pulses with the energy larger than $E_{\rm min}$ per
day, we can derive the actual observed rate on sky as
\begin{equation}
\mathcal{R} = \int_0^{0.5} \dd z \frac{\dd V}{\dd z} \eta \zeta r({>} E_{min}) \frac{R_0 (1 + z)^{3.28}}{(1 + z)} \tau,
\end{equation}
where $\eta$ is the active duty cycle, $\zeta$ is the beaming effect
and $1/(1 + z)$ is the effect of the time dilation. According the
observation of FRB 121102
\citep{Chatterjee2017Natur.541...58C,Zhang2018ApJ...866..149Z}, the
repeating FRBs are not always active. We introduce $\eta$ to
represent the proportion of activate period in the total life and
its value is taken as $\eta = 0.3$. A typical value $\zeta = 0.1$ is
taken for beaming effect \citep{2017ApJ...843...84N}. We take the
cumulative distribution of energy for VLA at 3 GHz as $r(>E_{\rm
min})$. The minimum energy in $r(>E_{\rm min})$ can be calculated as
$E_{\rm min} = 4 \pi d_L^2 F_{\rm min} \Delta \nu$, where $F_{\rm
min}=0.5$Jy ms is the fluence limit
\citep{Scholz2016ApJ...833..177S}. For the observed FRB rate,
\citet{Cao2018ApJ...858...89C} estimated $\mathcal{R}>F_{\rm
min}\sim 1.4\times 10^4$ events per day. Using above information, we
obtain $R_0 \tau \sim 2\times10^3\, \rm{Gpc}^{-3}$. The volume
density of repeating FRB sources averaged over $0 < z < 0.5$ is
$R_{\rm FRB} \tau \sim 5\times10^3\, \rm{Gpc}^{-3}$.
\citet{Lu2016MNRAS.461L.122L} derived the volume density of
repeating FRBs as $\sim 10^2-10^4 \rm{Gpc}^{-3}$, which agrees well
with ours. Assuming the luminosity function of repeating FRBs,
\citet{2017ApJ...843...84N} estimated $R_{\rm FRB}\tau \sim 10^4
\rm{Gpc}^{-3}$. Considering large uncertainty of FRB rate, ours is
marginally consistent their result. Because the possible progenitors
of superluminous supernovae (SLSNe), long gamma-ray bursts (LGRBs),
short gamma-ray bursts (SGRBs) and repeating FRBs are magnetars
\citep{Metzger17}, we compare their volume rates. For superluminous
supernovae, the average rate at $0<z<0.5$ is about $<R_{SLSN}>\sim
40$ Gpc$^{-3}$yr$^{-1}$ \citep{Quimby13}.
\citet{Yu2015ApJS..218...13Y} estimated the rate of LGRB is about
$R_{\rm LGRB}=7.2(1+z)^{0.04}$ Gpc$^{-3}$yr$^{-1}$. Averaging over
$0 < z < 0.5$ and taking their beaming factor of 20
\citep{Fong2015ApJ...815..102F}, we find $<R_{\rm LGRB}>\approx 180$
Gpc$^{-3}$yr$^{-1}$. Considering the beaming effect,
\citet{Zhang2018ApJ...852....1Z} derived the volume rate of SGRB is
$<R_{\rm SGRB}>\approx 200$ Gpc$^{-3}$yr$^{-1}$. If SLSNe and GRBs
are the progenitors of millisecond magnetars that power FRBs,
observed FRB rates require a lifetime of $\tau\sim 10-100$ years,
which is consistent with the estimations of Metzger et al. (2017)
and Cao et al. (2017).

Second, we compare our result with the energy distribution for
non-repeating FRBs. \citet{Lu2016MNRAS.461L.122L} assumed FRBs are
produced by neutron stars at cosmological distances and its rate
tracks star formation rate. They found that the observations of
non-repeating FRBs are consistent with a universal energy
distribution with power-law index $1.5<\alpha_E<2.2$, which is
consistent with our result. Moreover, the value of $\alpha_E$ has a
relative small range in our paper. More recently,
\citet{Zhang2019arXiv190601176} found the value of
$1.6<\alpha_E<2.0$ from the cumulative redshift distribution of
ASKAP and Parkes samples if the formation rate of FRB has a time
delay (3-5 Gyr) relative to cosmic star formation rate. However, if
the formation rate of FRB is proportional to the star formation
rate, the value of $\alpha_E$ is 2.3. So our results support that
the central magnetar is formed by merger of binary neutron stars,
which is consistent with the large offsets relative to the hist
centers of FRB 180924 \citet{Bannister19} and FRB 190523
\citet{Ravi19}. \citet{Lu2019arXiv190300014L} found the value of
$\alpha_E\sim 1.7$ from the dispersion measurement distribution of
ASKAP FRB sample. This value is dramatically consistent with our
result. The similar energy distributions between repeating FRB
121102 and non-repeating FRBs may indicate that they share the same
underlying physical mechanism \citep{Lu2019arXiv190300014L}.

Third, the energy distributions of other related phenomena are
discussed. The giant pulses of Crab show a steeper energy
distribution with $\alpha_E=2.1-3.5$ \citep{Mickaliger2012ApJ...760...64M}. FRB
energy distribution is consistent with magnetar burst ($\alpha_E\sim
1.6$; \citet{Prieskorn2012ApJ...755....1P,Wang2017}), type I X-ray
bursts ($\alpha_E\sim 1.5-1.7$; \citet{Wang2017MNRAS.471.2517W}) and other
avalanche events from self-organized criticality systems \citep{Katz1986,
  Bak1987PhRvL..59..381B,Lu1991ApJ...380L..89L,Aschwanden2011soca.book.....A,
  Wang2013NatPh...9..465W,Zhang2019arXiv190311895Z}.

\acknowledgements

We thank the anonymous referee for helpful comments. This work is supported by the National Natural Science Foundation of
China (grant No. U1831207).


\begin{deluxetable}{ccccc}
    \tablecaption{The best-fitting results in the high-energy cutoff case.\label{tab:results}}
    \tablehead{\colhead{Telescope} & \colhead{$ \alpha_E $} & \colhead{$ E\_{max} (10^{37}$erg)} & \colhead{A}  & \colhead{Reference}}
    \startdata
    VLA at 3 GHz        & 1.63$\pm0.19$ & -             & 1.30$\pm0.37\times 10^{25}$       & \citet{Chatterjee2017Natur.541...58C}                         \\
    Arecibo at 1.4 GHz  & 1.83$\pm0.09$ & 21$\pm17$     & 2.60$\pm0.12\times 10^{32}$       & \citet{Gourdji2019arXiv190302249G}                            \\
    Arecibo at 4.5 GHz  & 1.72$\pm0.02$ & 248$\pm48$    & 3.01$\pm0.25\times 10^{28}$       & \citet{Michilli2018Natur.553..182M}                           \\
    GBT at 2 GHz        & 1.67$\pm0.07$ & 316$\pm93$    & 4.58$\pm0.48\times 10^{26}$       & \citet{Scholz2016ApJ...833..177S,Scholz2017ApJ...846...80S}   \\
    Arecibo at 1.4 GHz  & 1.63$\pm0.21$ & 204$\pm90$    & 4.33$\pm0.87\times 10^{24}$       & \citet{Spitler2016Natur.531..202S}                            \\
    GBT at 4-8 GHz      & 1.56$\pm0.02$ & 307$\pm34$    & 9.78$\pm0.21\times 10^{22}$       & \citet{Zhang2018ApJ...866..149Z}
    \enddata
\end{deluxetable}

\begin{deluxetable}{ccccc}
    \tablecaption{The best-fitting results in the low-energy threshold and high-energy cutoff case, and the low-energy threshold case.\label{tab:Otherresults}}
    \tablehead{\colhead{Telescope} & \colhead{$ \alpha_E $\tablenotemark{$a$}} &   \colhead{$ \alpha_E $\tablenotemark{$b$}} & \colhead{$ E_{cut} $($ \times 10^{37} $erg)}&\colhead{Reference}}
    \startdata
    VLA at 3 GHz        & 1.74$\pm0.03$ & 1.73$\pm0.02$     &   2.31        &   \citet{Chatterjee2017Natur.541...58C}                       \\
    Arecibo at 1.4 GHz  & 2.34$\pm0.07$ & 2.34$\pm0.11$     &   2.00        &   \citet{Gourdji2019arXiv190302249G}                          \\
    Arecibo at 4.5 GHz  & 1.74$\pm0.03$ & 1.72$\pm0.02$     &   10.24       &   \citet{Michilli2018Natur.553..182M}                         \\
    GBT at 2 GHz        & 1.76$\pm0.04$ & 1.65$\pm0.07$     &   4.68        &   \citet{Scholz2016ApJ...833..177S,Scholz2017ApJ...846...80S} \\
    Arecibo at 1.4 GHz  & 1.76$\pm0.15$ & 1.63$\pm0.21$     &   2.00        &   \citet{Spitler2016Natur.531..202S}                          \\
    GBT at 4-8 GHz      & 1.69$\pm0.01$ & 1.61$\pm0.02$     &   5.27        &   \citet{Zhang2018ApJ...866..149Z}
    \enddata
    \tablenotetext{a}{Power-law indices for the low-energy threshold case.}
    \tablenotetext{b}{Power-law indices for the low-energy threshold and high-energy cutoff case.}
\end{deluxetable}

\begin{deluxetable}{ccccccc}
 \tabletypesize{\footnotesize}
  \tablecaption{The VLA observation at 3 GHz from Chatterjee et al.(2017). \label{tab:law}}
  \tablehead{ \colhead{MJD} & \colhead{Fluence (Jy ms)} & \colhead{Central Frequency (GHz)}
    & \colhead{Bandwidth (MHz)} & \colhead{Width (ms)}
    & \colhead{Energy ($10^{38}$ erg)} }
  \startdata
        57623.74402 & -               & 2.80            & 290                     & 2.00       & 12                                            \\
        57633.67986 & -               & 3.20            & 510                     & 2.05       & 98                                               \\
        57633.69516 & -               & \textless{}2.50 & \textless{}290          & 2.50       & 7                                              \\
        57638.49937 & -               & 3.10            & 420                     & 1.30       & 3                                               \\
        57643.45730 & -               & 2.80            & 510                     & 1.90       & 34                                             \\
        57645.42959 & -               & 2.80            & 380                     & 1.10       & 4                                               \\
        57646.43691 & -               & \textless{}2.50 & \textless{}430          & 2.50       & 10                                              \\
        57648.43691 & -               & 2.80            & 470                     & 1.40       & 7                                             \\
        57649.45176 & -               & 3.00            & 690                     & 2.10       & 12
  \enddata
\end{deluxetable}

\begin{deluxetable}{ccccccc}
 \tabletypesize{\footnotesize}
  \tablecaption{The Arecibo observation at 1.4 GHz from Gourdji et al. (2019). \label{tab:gourdji}}
  \tablehead{ \colhead{MJD} & \colhead{Fluence (Jy ms)} & \colhead{Central Frequency (GHz)}
    & \colhead{Bandwidth (MHz)} & \colhead{Width (ms)}
    & \colhead{Energy ($10^{38}$ erg)}}
  \startdata
        57644.41107 & 0.80            & 1.40              & 237       & 1.99      & 2.14         \\
        57644.41412 & 0.11            & 1.64              & 176       & 5.40      & 0.22          \\
        57644.41488 & 0.09            & 1.54              & 244       & 2.60      & 0.25          \\
        57644.41631 & 0.14            & 1.67              & 115       & 4.20      & 0.18         \\
        57644.43017 & 0.09            & 1.68              & 104       & 2.40      & 0.11       \\
        57644.43017 & 0.16            & 1.65              & 151       & 4.40      & 0.27         \\
        57644.43224 & 0.11            & 1.69              & 90        & 1.50      & 0.11          \\
        57644.43879 & 0.19            & 1.49              & 218       & 5.10      & 0.47           \\
        57644.43884 & 0.15            & 1.40              & 152       & 5.60      & 0.26          \\
        57644.44359 & 0.07            & 1.41              & 211       & 2.10      & 0.17            \\
        57644.44679 & 0.03            & 1.57              & 324       & 0.73      & 0.11           \\
        57644.44773 & 0.40            & 1.38              & 55        & 6.00      & 0.25          \\
        57644.44991 & 0.07            & 1.66              & 138       & 2.00      & 0.11         \\
        57644.45160 & 0.22            & 1.67              & 128       & 3.30      & 0.32          \\
        57644.45448 & 0.60            & 1.36              & 132       & 9.10      & 0.89           \\
        57644.45788 & 0.03            & 1.40              & 239       & 1.10      & 0.08          \\
        57644.46622 & 0.20            & 1.66              & 131       & 4.20      & 0.30         \\
        57644.46809 & 0.21            & 1.66              & 133       & 7.70      & 0.31           \\
        57645.41109 & 0.20            & 1.53              & 392       & 1.78      & 0.88           \\
        57645.41165 & 0.17            & 1.45              & 144       & 3.70      & 0.28           \\
        57645.41364 & 0.13            & 1.60              & 268       & 4.30      & 0.39         \\
        57645.41747 & 0.08            & 1.43              & 177       & 4.70      & 0.16            \\
        57645.41790 & 0.09            & 1.65              & 168       & 2.40      & 0.17            \\
        57645.42026 & 0.24            & 1.37              & 194       & 13.50     & 0.52           \\
        57645.42245 & 0.13            & 1.62              & 215       & 3.80      & 0.32           \\
        57645.42414 & 0.08            & 1.60              & 255       & 4.00      & 0.23          \\
        57645.42890 & 0.14            & 1.49              & 280       & 8.20      & 0.44        \\
        57645.43062 & 0.09            & 1.63              & 208       & 2.80      & 0.21           \\
        57645.43148 & 0.09            & 1.66              & 149       & 1.90      & 0.15           \\
        57645.44081 & 0.22            & 1.67              & 117       & 3.00      & 0.29           \\
        57645.44448 & 0.10            & 1.66              & 139       & 2.10      & 0.16           \\
        57645.44492 & 0.25            & 1.46              & 134       & 6.10      & 0.38            \\
        57645.44764 & 0.17            & 1.39              & 110       & 4.00      & 0.21           \\
        57645.44880 & 0.14            & 1.47              & 196       & 1.47      & 0.31           \\
        57645.44999 & 0.24            & 1.43              & 223       & 9.20      & 0.60           \\
        57645.44999 & 0.12            & 1.65              & 151       & 2.40      & 0.20           \\
        57645.45343 & 0.20            & 1.44              & 142       & 2.80      & 0.32            \\
        57645.45364 & 0.30            & 1.67              & 119       & 6.20      & 0.40           \\
        57645.46211 & 0.27            & 1.38              & 134       & 7.00      & 0.41            \\
        57645.46419 & 0.09            & 1.62              & 226       & 3.70      & 0.23           \\
        57645.47445 & 0.17            & 1.67              & 130       & 4.80      & 0.25
  \enddata
\end{deluxetable}

\begin{deluxetable}{ccccccc}
 \tabletypesize{\footnotesize}
  \tablecaption{The Arecibo observation at 4.5 GHz from Michilli et al. (2018). \label{tab:michilli}}
  \tablehead{ \colhead{MJD} & \colhead{Fluence (Jy ms)} & \colhead{Central Frequency (GHz)}
    & \colhead{Bandwidth (MHz)} & \colhead{Width (ms)}
    & \colhead{Energy ($10^{38}$ erg)} }
  \startdata
        57747.12956 & 0.70           & 4.50                   & 800            & 0.80      & 6.31          \\
        57747.13719 & 0.20           & 4.50                   & 800            & 0.85      & 1.80          \\
        57747.14627 & 0.20           & 4.50                   & 800            & 0.22      & 1.80          \\
        57747.15157 & 0.09           & 4.50                   & 800            & 0.55      & 0.81          \\
        57747.15447 & 0.10           & 4.50                   & 800            & 0.76      & 0.90          \\
        57747.16029 & 0.05           & 4.50                   & 800            & 0.03      & 0.45          \\
        57747.16034 & 0.20           & 4.50                   & 800            & 0.31      & 1.80          \\
        57747.16583 & 0.50           & 4.50                   & 800            & 1.36      & 4.51          \\
        57747.16637 & 0.30           & 4.50                   & 800            & 1.92      & 2.70          \\
        57747.17597 & 0.20           & 4.50                   & 800            & 0.98      & 1.80          \\
        57748.12564 & 0.10           & 4.50                   & 800            & 0.95      & 0.90          \\
        57748.15352 & 0.20           & 4.50                   & 800            & 0.42      & 1.80          \\
        57748.15521 & 0.60           & 4.50                   & 800            & 0.78      & 5.41          \\
        57748.15761 & 0.20           & 4.50                   & 800            & 0.15      & 1.80          \\
        57748.17570 & 0.40           & 4.50                   & 800            & 0.54      & 3.61          \\
        57772.12903 & 0.60           & 4.50                   & 800            & 0.74      & 5.41
  \enddata
\end{deluxetable}

\begin{deluxetable}{ccccccc}
 \tabletypesize{\footnotesize}
  \tablecaption{The GBT observation at 2 GHz from Scholz et al. (2016). \label{ref:scholz}}
  \tablehead{ \colhead{MJD} & \colhead{Fluence (Jy ms)} & \colhead{Central Frequency (GHz)}
    & \colhead{Bandwidth (MHz)} & \colhead{Width (ms)}
    & \colhead{Energy ($10^{38}$ erg)}  }
  \startdata
        57339.35605 & 0.20           & 2.00                   & 800            & 6.73      & 1.80          \\
        57345.44769 & 0.40           & 2.00                   & 800            & 6.10      & 3.61          \\
        57345.45249 & 0.20           & 2.00                   & 800            & 6.14      & 1.80          \\
        57345.45760 & 0.08           & 2.00                   & 800            & 4.30      & 0.72          \\
        57345.46241 & 0.60           & 2.00                   & 800            & 5.97      & 5.41          \\
        57647.23235 & 0.82           & 2.00                   & 800            & 2.16      & 7.39          \\
        57647.23235 & 0.16           & 2.00                   & 800            & 1.94      & 1.44          \\
        57649.17381 & 1.32           & 2.00                   & 800            & 3.45      & 11.90         \\
        57649.21821 & 0.34           & 2.00                   & 800            & 0.88      & 3.07          \\
        57765.04953 & 0.33           & 2.00                   & 800            & 1.40      & 2.98          \\
        57765.06479 & 0.83           & 2.00                   & 800            & 1.79      & 7.48          \\
        57765.06905 & 0.62           & 2.00                   & 800            & 2.97      & 5.59          \\
        57765.10083 & 0.18           & 2.00                   & 800            & 2.46      & 1.62          \\
        57765.12078 & 1.08           & 2.00                   & 800            & 1.36      & 9.74          \\
        57765.13650 & 0.22           & 2.00                   & 800            & 1.68      & 1.98
  \enddata
\end{deluxetable}

\begin{deluxetable}{ccccccc}
 \tabletypesize{\footnotesize}
  \tablecaption{The Arecibo observation at 1.4 GHz from Spilter et al. (2016). \label{tab:spilter}}
  \tablehead{ \colhead{MJD} & \colhead{Fluence (Jy ms)} & \colhead{Central Frequency (GHz)}
    & \colhead{Bandwidth (MHz)} & \colhead{Width (ms)}
    & \colhead{Energy ($10^{38}$ erg)} }
  \startdata
        56233.28284 & 0.10           & 1.40                   & 322             & 3.30      & 0.36          \\
        57159.73760 & 0.10           & 1.40                   & 322             & 3.80      & 0.36          \\
        57159.74422 & 0.10           & 1.40                   & 322             & 3.30      & 0.36          \\
        57175.69314 & 0.20           & 1.40                   & 322             & 4.60      & 0.73          \\
        57175.69973 & 0.09           & 1.40                   & 322             & 8.70      & 0.33          \\
        57175.74258 & 0.06           & 1.40                   & 322             & 2.80      & 0.22          \\
        57175.74284 & 0.06           & 1.40                   & 322             & 6.10      & 0.22          \\
        57175.74351 & 0.90           & 1.40                   & 322             & 6.60      & 3.27          \\
        57175.74567 & 0.30           & 1.40                   & 322             & 6.00      & 1.09          \\
        57175.74762 & 0.20           & 1.40                   & 322             & 8.00      & 0.73          \\
        57175.74829 & 1.00           & 1.40                   & 322             & 3.06      & 3.63
  \enddata
\end{deluxetable}
\clearpage

\startlongtable
\begin{deluxetable}{ccccccc}
 \tabletypesize{\footnotesize}
  \tablecaption{The GBT observation at 4-8 GHz from Zhang et al. (2018). \label{tab:zhang}}
  \tablehead{ \colhead{MJD} & \colhead{Fluence (Jy ms)} & \colhead{Central Frequency (GHz)}
    & \colhead{Bandwidth (MHz)} & \colhead{Width (ms)}
    & \colhead{Energy ($10^{38}$ erg)} }
  \startdata
        57991.4099  & 0.61              & 6.05              & 3900           & 1.43      & 26.64        \\
        57991.40993 & 0.03              & 6.75              & 2300           & 2.15      & 0.66         \\
        57991.41002 & 0.05              & 6.3               & 2400           & 1.43      & 1.37           \\
        57991.41007 & 0.09              & 6.25              & 2500           & 2.51      & 2.55         \\
        57991.41156 & 0.05              & 6.25              & 2500           & 1.79      & 1.49         \\
        57991.41209 & 0.04              & 5.7               & 1400           & 1.79      & 0.69         \\
        57991.41231 & 0.03              & 6.3               & 2400           & 1.08      & 0.71         \\
        57991.41276 & 0.07              & 5.95              & 3900           & 1.43      & 2.92         \\
        57991.41276 & 0.07              & 5.45              & 2900           & 1.79      & 2.20         \\
        57991.41293 & 0.15              & 6                 & 3000           & 3.58      & 4.94        \\
        57991.41302 & 0.06              & 7                 & 1800           & 0.72      & 1.15        \\
        57991.41336 & 0.05              & 7                 & 1800           & 1.43      & 0.95         \\
        57991.41346 & 0.36              & 6                 & 4000           & 1.08      & 16.37        \\
        57991.41371 & 0.26              & 6                 & 4000           & 1.08      & 11.85        \\
        57991.41384 & 0.26              & 6                 & 4000           & 1.43      & 11.93        \\
        57991.41613 & 0.04              & 6.05              & 2900           & 1.79      & 1.15         \\
        57991.41619 & 0.05              & 6.3               & 2400           & 1.79      & 1.25         \\
        57991.41621 & 0.02              & 5.5               & 2000           & 1.08      & 0.52         \\
        57991.41644 & 0.04              & 5.95              & 3900           & 1.08      & 1.59         \\
        57991.41663 & 0.35              & 6.05              & 3900           & 1.08      & 15.35        \\
        57991.41727 & 0.05              & 6.25              & 2500           & 1.08      & 1.48         \\
        57991.41738 & 0.03              & 5.95              & 1900           & 0.72      & 0.63         \\
        57991.41771 & 0.04              & 6.25              & 2500           & 1.08      & 0.99         \\
        57991.41787 & 0.08              & 6.05              & 3900           & 1.08      & 3.56         \\
        57991.41863 & 0.1               & 6.05              & 3900           & 1.43      & 4.30         \\
        57991.41903 & 0.04              & 6                 & 2000           & 1.43      & 0.90          \\
        57991.41945 & 0.04              & 5.85              & 1700           & 1.43      & 0.85         \\
        57991.41945 & 0.02              & 7.1               & 1000           & 1.08      & 0.24         \\
        57991.41945 & 0.04              & 5.9               & 1600           & 1.08      & 0.65         \\
        57991.41946 & 0.03              & 5.5               & 2000           & 1.79      & 0.58         \\
        57991.41946 & 0.03              & 5.55              & 1900           & 1.43      & 0.63         \\
        57991.42087 & 0.03              & 7.15              & 1300           & 1.43      & 0.42         \\
        57991.42121 & 0.06              & 6                 & 4000           & 0.72      & 2.74          \\
        57991.42171 & 0.08              & 5.95              & 3900           & 0.72      & 3.45          \\
        57991.42212 & 0.07              & 5.5               & 2000           & 1.79      & 1.54         \\
        57991.42214 & 0.04              & 6.7               & 2400           & 1.79      & 1.07         \\
        57991.42294 & 0.13              & 6.05              & 3900           & 1.79      & 5.77         \\
        57991.42301 & 0.03              & 6                 & 4000           & 2.15      & 1.19         \\
        57991.42427 & 0.05              & 5.55              & 1900           & 1.79      & 1.13         \\
        57991.42454 & 0.02              & 5.05              & 900            & 1.08      & 0.18          \\
        57991.42592 & 0.03              & 5.55              & 900            & 2.15      & 0.29         \\
        57991.42639 & 0.07              & 6.05              & 1900           & 2.15      & 1.44         \\
        57991.42655 & 0.16              & 6.05              & 3900           & 0.72      & 6.88         \\
        57991.42839 & 0.07              & 6.7               & 2400           & 2.51      & 2.02         \\
        57991.42859 & 0.04              & 5.95              & 3900           & 1.08      & 1.59         \\
        57991.43043 & 0.05              & 5.95              & 3900           & 1.08      & 2.24         \\
        57991.4311  & 0.06              & 6.2               & 1400           & 3.23      & 0.91         \\
        57991.43167 & 0.04              & 6                 & 2000           & 1.79      & 0.93          \\
        57991.43197 & 0.16              & 5.95              & 3900           & 1.08      & 7.16         \\
        57991.43228 & 0.05              & 6                 & 2000           & 1.43      & 1.08         \\
        57991.43484 & 0.02              & 6.45              & 1900           & 1.79      & 0.52         \\
        57991.4352  & 0.06              & 6.55              & 1900           & 1.43      & 1.20         \\
        57991.43656 & 0.06              & 6.95              & 1900           & 1.08      & 1.21         \\
        57991.43785 & 0.02              & 6.7               & 2400           & 1.08      & 0.66         \\
        57991.43849 & 0.02              & 5.75              & 1500           & 1.08      & 0.38         \\
        57991.43936 & 0.23              & 6                 & 4000           & 1.79      & 10.56        \\
        57991.44621 & 0.02              & 6.2               & 1400           & 0.72      & 0.28         \\
        57991.44699 & 0.01              & 6.25              & 1500           & 0.72      & 0.22         \\
        57991.44793 & 0.05              & 5.95              & 2900           & 2.51      & 1.71         \\
        57991.44805 & 0.03              & 4.95              & 900            & 2.15      & 0.26          \\
        57991.44843 & 0.14              & 5.95              & 3900           & 1.08      & 6.04         \\
        57991.45018 & 0.02              & 5.3               & 1400           & 0.72      & 0.31         \\
        57991.45047 & 0.03              & 6.75              & 2300           & 1.43      & 0.89         \\
        57991.45304 & 0.07              & 5.7               & 2400           & 1.43      & 1.82         \\
        57991.45851 & 0.02              & 5.3               & 1400           & 1.79      & 0.34         \\
        57991.46087 & 0.03              & 7.2               & 1400           & 1.08      & 0.43         \\
        57991.46371 & 0.05              & 6.25              & 1900           & 1.08      & 1.01         \\
        57991.46394 & 0.04              & 5.5               & 2000           & 1.43      & 0.99         \\
        57991.46394 & 0.04              & 6                 & 3000           & 1.08      & 1.24         \\
        57991.47019 & 0.02              & 5.8               & 1400           & 0.72      & 0.32         \\
        57991.47115 & 0.03              & 6.25              & 1500           & 2.15      & 0.46         \\
        57991.47287 & 0.02              & 5.05              & 900            & 1.79      & 0.20         \\
        57991.48003 & 0.04              & 5.5               & 2000           & 1.79      & 0.79          \\
        57991.4886  & 0.03              & 6.45              & 900            & 2.87      & 0.27          \\
        57991.49316 & 0.02              & 5.45              & 1900           & 1.79      & 0.40         \\
        57991.49751 & 0.05              & 5.5               & 2000           & 1.43      & 1.13         \\
        57991.49839 & 0.02              & 5.55              & 900            & 2.15      & 0.25         \\
        57991.49882 & 0.05              & 5.7               & 2400           & 1.79      & 1.39         \\
        57991.51086 & 0.02              & 6.5               & 2000           & 0.72      & 0.54         \\
        57991.51311 & 0.1               & 5.5               & 2000           & 1.08      & 2.17         \\
        57991.51923 & 0.05              & 6.5               & 2000           & 2.87      & 1.12         \\
        57991.51923 & 0.07              & 6.5               & 2000           & 1.43      & 1.49         \\
        57991.5594  & 0.03              & 6.05              & 3900           & 0.72      & 1.31         \\
        57991.56282 & 0.06              & 6.5               & 2800           & 1.08      & 1.85         \\
        57991.56945 & 0.07              & 5.5               & 2000           & 2.87      & 1.57         \\
        57991.571   & 0.05              & 5.55              & 1900           & 1.43      & 1.05         \\
        57991.57444 & 0.02              & 6.45              & 900            & 2.51      & 0.18         \\
        57991.58815 & 0.05              & 5.5               & 2000           & 1.08      & 1.08         \\
        57991.59051 & 0.04              & 6.3               & 2400           & 1.08      & 1.14         \\
        57991.59211 & 0.05              & 6                 & 2000           & 1.08      & 1.17         \\
        57991.59543 & 0.02              & 6.25              & 1500           & 0.72      & 0.35         \\
        57991.59576 & 0.03              & 6.25              & 1500           & 1.43      & 0.49         \\
        57991.60226 & 0.02              & 5.15              & 1300           & 0.72      & 0.31
  \enddata
\end{deluxetable}

\end{document}